\documentclass[11pt]{article}
\usepackage{amsmath}
\usepackage{cite}
\usepackage{graphicx}
\usepackage{amsfonts}
\usepackage{amssymb}
\usepackage[left=0.75in,right=0.75in,top=1in,bottom=1in]{geometry}
\providecommand{\U}[1]{\protect\rule{.1in}{.1in}}

\csname @addtoreset\endcsname{equation}{section}
\newcommand{\eq}{\begin{equation}}
\newcommand{\feq}{\end{equation}}
\newcommand{\eqn}{\begin{eqnarray}}
\newcommand{\feqn}{\end{eqnarray}}
\newcommand{\arr}{\begin{eqnarray*}}
\newcommand{\farr}{\end{eqnarray*}}

\begin{document}
\begin{titlepage}
\vspace{0.3cm}
\begin{center}
\renewcommand{\thefootnote}{\fnsymbol{footnote}}
{\Large\bfseries 
Single-Variable Solutions in Supergravity}
\vskip2cm
{\large W.~A.~Sabra and R.~Slim}
\renewcommand{\thefootnote}{\arabic{footnote}}
\setcounter{footnote}{0}
\vskip1cm
{\small
Physics 
Department, American University of Beirut, Lebanon.}
\end{center}
\vskip2cm
\begin{center}
\textbf{Abstract}
\end{center}
 We construct four families of spacetime metrics depending on a single variable for a broad class of 
$D$-dimensional 
gravitational theories coupled to scalar and Abelian gauge fields. 
 As applications of the general formalism, we derive one-variable solutions of ungauged $\mathcal{N}=2$, $D=4$ supergravity coupled to vector multiplets. We also obtain explicit solutions for a consistent truncation of $\mathcal{N}=8$, $D=4$ supergravity, 
 as well as 
for theories whose scalar fields parametrize the symmetric coset manifolds 
 $SL(N,\mathbb{R})/SO(N,\mathbb{R})$.  
In all cases, the geometry of the scalar manifold plays a central role in determining the structure of the resulting 
solutions with nontrivial scalar and gauge field configurations.
\end{titlepage}
\section{Introduction}
In this paper, we investigate spacetime solutions that depend on a single
variable in gravitational theories coupled to scalar and Abelian gauge
fields. Our primary focus is on four-dimensional $\mathcal{N}=2$
supergravity theories coupled to vector multiplets \cite{superbook},
together with higher-dimensional gravitational theories whose scalar fields
parametrize the symmetric spaces $SL(N,\mathbb{R})/SO(N,\mathbb{R})$.

A prototypical example of a one-variable vacuum solution is the Kasner
metric \cite{kasner}, 
\begin{equation}
ds^{2}=x^{2a_{1}}dx^{2}+x^{2a_{2}}dy^{2}+x^{2a_{3}}dz^{2}+x^{2a_{4}}dw^{2}%
\text{ },  \label{o}
\end{equation}%
where the constant Kasner exponents satisfy 
\begin{align}
a_{2}+a_{3}+a_{4}& =1+a_{1}\text{ },\text{ }  \notag \\
a_{2}^{2}+a_{3}^{2}+a_{4}^{2}& =\left( 1+a_{1}\right) ^{2}.  \label{ig}
\end{align}%
After an appropriate analytic continuation, this yields the Lorentzian
metric 
\begin{equation}
ds^{2}=-dt^{2}+t^{2p_{1}}dx^{2}+t^{2p_{2}}dy^{2}+t^{2p_{3}}dz^{2}\text{ ,}
\label{km}
\end{equation}%
commonly referred to as the Kasner solution. This metric provides a
canonical realization of a Bianchi type I cosmology \cite{macel}. Kasner
spacetimes and their generalizations have been extensively studied as models
of anisotropic cosmologies and spacetime singularities. They also admit a
natural extension to arbitrary spacetime dimensions and signatures \cite%
{harvey}, 
\begin{equation}
ds^{2}=\epsilon _{0}d\tau ^{2}+\epsilon _{i}\tau ^{2p_{i}}\left(
dx^{i}\right) ^{2}\text{ },
\end{equation}%
where the Kasner exponents satisfy%
\begin{equation}
\sum_{i=1}^{D-1}p_{i}=\sum_{i=1}^{D-1}p_{i}^{2}=1\text{ },
\end{equation}%
and $\epsilon _{0}$ and $\epsilon _{i}$ are constants taking the values $\pm
1.$

A systematic analysis of one-variable vacuum solutions in four-dimensional
gravity was carried out in \cite{landau}, and extends naturally to arbitrary
dimensions and signatures. For metrics of the form 
\begin{equation}
ds^{2}=\epsilon _{0}d\tau ^{2}+g_{ij}(\tau )dx^{i}dx^{j}\text{ },
\end{equation}%
with $i,j=1,\ldots ,D-1,$ the Einstein equations reduce to 
\begin{align}
\ddot{\omega}& =0\text{ },  \notag \\
\kappa ^{i}{}_{j}\kappa ^{j}{}_{i}& =4\left( \frac{\dot{\omega}}{\omega }%
\right) ^{2}\text{ },  \notag \\
\frac{d}{d\tau }\left( \omega \kappa ^{i}{}_{j}\right) & =0\text{ },
\label{h2}
\end{align}%
where an overdot denotes differentiation with respect to $\tau ,$ $\kappa
^{i}{}_{j}=g^{ik}\dot{g}_{kj}$ and $\omega =\sqrt{\left\vert g\right\vert }$%
. These equations admit two inequivalent classes of solutions. The first
corresponds to $\omega =\tau $ and leads to the metric 
\begin{equation}
ds^{2}=\epsilon _{0}d\tau ^{2}+h_{ik}\left( e^{\left( 2\ln \tau \right)
\lambda }\right) ^{k}{}_{j}dx^{i}dx^{j}\text{ },  \label{met1}
\end{equation}%
where $h_{ij}$ are the components of a constant symmetric matrix and,
together with the constants $\lambda ^{i}{}_{j}$, satisfy 
\begin{align}
\lambda ^{i}{}_{j}\lambda ^{j}{}_{i}& =\lambda ^{i}{}_{i}=1\text{ },  \notag
\\
h_{ik}\lambda ^{k}{}_{j}& =h_{jk}\lambda ^{k}{}_{i}\text{ }.
\end{align}%
The second class corresponds to $\omega =1$ and was identified in \cite%
{harvey2}. These solutions were not included in the original analyses of 
\cite{kasner, landau} and correspond, in Kasner's notation, to the special
case $a_{1}=-1$. The metric takes the form 
\begin{equation}
ds^{2}=\epsilon _{0}d\chi ^{2}+h_{ik}\left( e^{\theta \chi }\right)
^{k}{}_{j}dx^{i}dx^{j}\text{ },
\end{equation}%
where $h_{ij}$ are again the components of a constant symmetric matrix and $%
\theta ^{i}{}_{j}$ are constants satisfying 
\begin{eqnarray}
\theta ^{i}{}_{i} &=&\theta ^{i}{}_{j}\theta ^{j}{}_{i}=0\text{ },  \notag \\
h_{ik}\theta ^{k}{}_{j} &=&h_{jk}\theta ^{k}{}_{i}\text{ }.
\end{eqnarray}%
In both classes of vacuum solutions, the values of $h_{ij}$ can be set to $%
\pm 1.$

Kasner-type solutions of four-dimensional gravity coupled to a scalar field
were investigated in \cite{Belinski}. Starting from the Kasner and
scalar-Kasner solutions, Melvin-type geometries \cite{melvin} were
constructed in \cite{dowker, kt} using solution-generating techniques
developed in \cite{har, earn}. More recently, generalized Kasner metrics and
other one-variable solutions have been obtained in several supergravity
theories \cite{s1, s2, s3, new, fs, 5d}.

The aim of the present work is to extend the analysis of one-variable vacuum
geometries to gravitational theories coupled to scalar and Abelian gauge
fields. Such theories naturally arise as bosonic sectors of supergravity and
string theory compactifications, where nontrivial scalar and gauge field
configurations play a fundamental role in determining the structure of exact
solutions. We develop a general framework for constructing one-variable
solutions in arbitrary spacetime dimensions and signatures and derive four
distinct families of solutions corresponding to different gauge field
configurations. We then apply this formalism to ungauged $\mathcal{N}=2$, $%
D=4$ supergravity coupled to vector multiplets, as well as to theories whose
scalar manifolds are the symmetric spaces $SL(N,\mathbb{R})/SO(N,\mathbb{R})$%
. In both cases, we obtain solutions supported by nontrivial scalar and
gauge field configurations.

The paper is organized as follows. In Section 2, we analyze the equations of
motion governing one-variable solutions of $D$-dimensional gravity coupled
to scalar and Abelian gauge fields. In Section 3, we apply the formalism to $%
\mathcal{N}=2$, $D=4$ supergravity and construct four families of solutions,
including explicit examples arising from a consistent truncation of $%
\mathcal{N}=8$, $D=4$ supergravity. Section 4, presents explicit solutions
for theories whose scalar fields parametrize the coset spaces $SL(N,\mathbb{R%
})/SO(N,\mathbb{R})$. Finally, Section 5 summarizes our results and
discusses possible directions for future research.

\section{One-variable solutions of gravity with gauge and scalar fields}

In this section, we derive one-variable solutions of $D$-dimensional
gravitational theories coupled to scalar and Abelian gauge fields for
arbitrary spacetime signatures. Throughout, we assume that all fields depend
on a single coordinate. We consider the general Lagrangian 
\begin{equation}
\mathcal{L}_{D}=\omega \!\!\left( R-2\mathcal{G}_{A\bar{B}}\partial _{\mu
}\phi ^{A}\partial ^{\mu }\bar{\phi}^{B}-{\frac{\varepsilon }{2}}G_{IJ}%
\mathcal{F}_{\mu \nu }^{I}\mathcal{F}^{J\mu \nu \,}\right) \text{ },
\label{act}
\end{equation}%
where $\mathcal{G}_{A\bar{B}}$ and $G_{IJ}$ denote the scalar and gauge
coupling metrics, respectively. In general, both metrics depend on the
scalar fields $\phi ^{A}$, and $\varepsilon =\pm 1.$ The Lagrangian (\ref%
{act}) encompasses the bosonic sectors of a broad class of supergravity
theories coupled to scalar and Abelian gauge fields. The geometry of the
scalar manifold is determined by the specific theory under consideration.
The corresponding Einstein equations are 
\begin{equation}
R_{\mu \nu }=2\mathcal{G}_{A\bar{B}}\partial _{\mu }\phi ^{A}\partial _{\nu }%
\bar{\phi}^{B}+\varepsilon G_{IJ}\left( \mathcal{F}_{\mu \lambda }^{I}%
\mathcal{F}_{\nu }^{J}{}^{\lambda }-{\frac{1}{2(D-2)}}g_{\mu \nu }\mathcal{F}%
_{\rho \sigma }^{I}\mathcal{F}^{J\rho \sigma }\right) \text{ ,}  \label{ein}
\end{equation}%
while the Maxwell equations are 
\begin{equation}
\partial _{\mu }(\omega G_{IJ}\mathcal{F}^{J\mu \nu })=0\text{ }.
\end{equation}%
The scalar field equations take the form 
\begin{equation}
{\frac{1}{\omega }}\partial _{\mu }(\omega g^{\mu \nu }\mathcal{G}_{A\bar{B}%
}\partial _{\nu }\bar{\phi}^{B})-\partial _{A}\mathcal{G}_{C\bar{D}}\partial
_{\mu }\phi ^{C}\partial ^{\mu }\bar{\phi}^{D}-{\frac{\varepsilon }{4}}%
\partial _{A}G_{JK}\mathcal{F}_{\mu \nu }^{J}\mathcal{F}^{K\mu \nu }=0\text{
. }
\end{equation}

Depending on the gauge field configuration, the resulting one-variable
solutions naturally split into two inequivalent classes. These are discussed
in the following subsections.

\subsection{Class 1 solutions}

The first class of solutions is described by the metric 
\begin{equation}
ds^{2}=\epsilon _{0}d\tau ^{2}+g_{ij}(\tau )dx^{i}dx^{j}=\epsilon _{0}d\tau
^{2}+g_{zz}(\tau )dz^{2}+g_{ab}(\tau )dx^{a}dx^{b}\text{ },
\end{equation}%
together with gauge field strengths satisfying 
\begin{equation}
G_{IJ}\mathcal{F}^{I\tau z}=\frac{q_{J}}{\omega }\text{ },
\end{equation}%
where $q_{J}$ are constants and $a,b=1,\ldots,D-2$. Substituting this ansatz
into the Einstein equations yields 
\begin{align}
\dot{g}_{zz}& =\frac{1}{\omega }\left( \theta ^{z}{}_{z}-2\left( D-3\right) 
\dot{\omega}\right) g_{zz}\text{ },  \notag \\
\dot{g}_{ab}& =\frac{1}{\omega }\left( g_{ac}\theta ^{c}{}_{b}+2\dot{\omega}%
g_{ab}\right) \text{ },  \label{metrictwo}
\end{align}%
where the matrices $\theta ^{i}{}_{j}$ arise as integration constants and
satisfy the tracelessness condition 
\begin{equation}
\theta ^{i}{}_{i}=\theta ^{z}{}_{z}+\theta ^{a}{}_{a}=0\text{ }.  \label{fa}
\end{equation}%
The remaining Einstein equations reduce to 
\begin{align}
2\omega ^{2}\mathcal{G}_{A\bar{B}}\partial _{\tau }\phi ^{A}\partial _{\tau }%
\bar{\phi}^{B}& =-\frac{1}{4}\theta ^{i}{}_{j}\theta ^{j}{}_{i}-\left(
D-2\right) \left[ \left( D-3\right) \dot{\omega}^{2}-\theta ^{z}{}_{z}\dot{%
\omega}+\omega \ddot{\omega}\right] \text{ },  \notag \\
\ddot{\omega}& =\frac{1}{\left( D-2\right) \omega }\varepsilon
g_{zz}G^{IJ}q_{J}q_{I}\text{ },  \label{g}
\end{align}%
while the scalar field equations become 
\begin{equation}
\partial _{\tau }(\omega \mathcal{G}_{A\bar{B}}\partial _{\tau }\bar{\phi}%
^{B})-\omega \partial _{A}\mathcal{G}_{C\bar{D}}\partial _{\tau }\phi
^{C}\partial _{\tau }\bar{\phi}^{D}=-\frac{{\varepsilon }}{2}\partial
_{A}G^{IJ}g_{zz}\frac{q_{I}q_{J}}{\omega }\text{ }.  \label{pi}
\end{equation}

The metric equations (\ref{metrictwo}) can be integrated independently of
the scalar manifold geometry. Introducing a new coordinate $\sigma $ through 
\begin{equation}
d\tau =Hd\sigma \text{ },\text{ \ \ \ \ \ \ \ \ }\omega =\sigma H\text{ },
\label{cs}
\end{equation}%
and defining 
\begin{equation}
p=\frac{1}{2}\theta ^{z}{}_{z}-\left( D-3\right) \text{ },\text{ \ \ }%
\lambda ^{a}{}_{b}=\frac{1}{2}\theta ^{a}{}_{b}+\delta ^{a}{}_{b}\text{ },
\label{bn}
\end{equation}%
the metric takes the form 
\begin{equation}
ds^{2}=H^{2}\left( \epsilon _{0}d\sigma ^{2}+h_{ac}\left( e^{\left( 2\log
\sigma \right) \lambda }\right) ^{c}{}_{b}dx^{a}dx^{b}\right) +h_{zz}\sigma
^{2p}H^{-2\left( D-3\right) }dz^{2}\text{ }.  \label{f1}
\end{equation}%
subject to the conditions 
\begin{equation}
p+\lambda ^{a}{}_{a}=1\text{ , \ \ \ }h_{ac}\lambda ^{c}{}_{b}=h_{bc}\lambda
^{c}{}_{a}\text{ }.  \label{c1}
\end{equation}

A second family of solutions is obtained by introducing the coordinate $\chi 
$ according to%
\begin{equation}
d\tau =\omega d\chi \text{ }.
\end{equation}%
The metric then takes the form

\begin{equation}
ds^{2}=\omega ^{2}\left( \epsilon _{0}d\chi ^{2}+h_{ac}\left( e^{\theta \chi
}\right) ^{c}{}_{b}dx^{a}dx^{b}\right) +h_{zz}e^{\mu \chi }\omega ^{-2\left(
D-3\right) }dz^{2}\text{ }.  \label{chad}
\end{equation}%
with the conditions%
\begin{equation}
\mu +\theta ^{a}{}_{a}=0\text{ , \ \ \ }h_{ac}\theta ^{c}{}_{b}=h_{bc}\theta
^{c}{}_{a}\text{ }.  \label{c2}
\end{equation}

The explicit form of these solutions follows from solving Eqs. (\ref{g}),
and (\ref{pi}). Consequently, the resulting spacetime geometries are
determined by the geometry of the scalar manifold and the associated gauge
coupling functions.

\subsection{Class 2 solutions}

The second class of one-variable solutions is characterized by the metric%
\begin{equation}
ds^{2}=\epsilon _{0}d\tau ^{2}+g_{ij}(\tau )dx^{i}dx^{j}=\epsilon _{0}d\tau
^{2}+g_{xx}(\tau )\left( dx\right) ^{2}+g_{yy}(\tau )\left( dy\right)
^{2}+g_{ab}(\tau )dx^{a}dx^{b}\text{ },
\end{equation}%
together with gauge field strengths satisfying%
\begin{equation}
\mathcal{F}_{xy}^{I}=p^{I}\text{ },  \label{gog}
\end{equation}%
where $p^{I}$ are constants and $a,b=1,\ldots ,D-3$. Substituting this
ansatz into the Einstein equations yields 
\begin{align}
\dot{g}_{xx}& =\frac{1}{\omega }\left( \theta ^{x}{}_{x}+2\dot{\omega}%
\right) g_{xx}\text{ },  \notag \\
\dot{g}_{yy}& =\frac{1}{\omega }\left( \theta ^{y}{}_{y}+2\dot{\omega}%
\right) g_{yy}\text{ },  \notag \\
\dot{g}_{ab}& =\frac{1}{\omega }\left( g_{ac}\theta ^{c}{}_{b}-\frac{2\dot{%
\omega}}{D-3}g_{ab}\right) \text{ }.
\end{align}%
The integration constants satisfy the tracelessness condition 
\begin{equation}
\theta ^{i}{}_{i}=\theta ^{x}{}_{x}+\theta ^{y}{}_{y}+\theta ^{a}{}_{a}=0%
\text{ }.
\end{equation}%
The remaining Einstein equations become 
\begin{align}
G_{IJ}p^{J}p^{I}g^{xx}g^{yy}& =-\varepsilon \epsilon _{0}\frac{(D-2)}{\left(
D-3\right) }\frac{\ddot{\omega}}{\omega }\text{ ,}  \notag \\
2\mathcal{G}_{A\bar{B}}\partial _{\tau }\phi ^{A}\partial _{\tau }\bar{\phi}%
^{B}\omega ^{2}& =-\frac{1}{4}\theta ^{i}{}_{j}\theta ^{j}{}_{i}+\left( 
\frac{D-2}{D-3}\right) \left( \theta ^{a}{}_{a}\dot{\omega}-\dot{\omega}^{2}-%
\ddot{\omega}\omega \right) \text{ },  \label{cs2}
\end{align}%
while the scalar field equations reduce to 
\begin{equation}
{\frac{1}{\omega }}\partial _{\tau }(\omega \mathcal{G}_{A\bar{B}}\partial
_{\tau }\bar{\phi}^{B})-\partial _{A}\mathcal{G}_{C\bar{D}}\partial _{\tau
}\phi ^{C}\partial _{\tau }\bar{\phi}^{D}-\frac{{\varepsilon }}{2}\epsilon
_{0}g^{yy}g^{xx}\partial _{A}G_{JK}q^{J}q^{K}=0\text{ . }  \label{sq}
\end{equation}%
Using the coordinate transformation (\ref{cs}), the metric becomes 
\begin{equation}
ds^{2}=H^{2}\left( \epsilon _{0}d\sigma ^{2}+h_{xx}\sigma
^{2n_{x}}dx^{2}+h_{yy}\sigma ^{2n_{y}}dy^{2}\right) +h_{ac}\left( e^{\left(
2\log \sigma \right) \lambda }\right) ^{c}{}_{b}H^{-\frac{2}{D-3}%
}dx^{a}dx^{b}  \label{genp}
\end{equation}%
where 
\begin{equation}
n_{x}=\frac{1}{2}\theta ^{x}{}_{x}+1,\ \ \ n_{y}=\frac{1}{2}\theta
^{y}{}_{y}+1,\text{ \ \ \ }\lambda ^{a}{}_{b}=\frac{1}{2}\theta ^{a}{}_{b}-%
\frac{1}{D-3}\delta ^{a}{}_{b}\text{ }.
\end{equation}%
The parameters satisfy 
\begin{eqnarray}
n_{x}+n_{y}+\lambda ^{a}{}_{a} &=&1\text{ },  \notag \\
h_{ac}\lambda ^{c}{}_{b} &=&h_{bc}\lambda ^{c}{}_{a}\text{ }.  \label{ab}
\end{eqnarray}

A second family of solutions is obtained by introducing the coordinate $\chi 
$, yielding the metric%
\begin{equation}
ds^{2}=\omega ^{2}\left( \epsilon _{0}d\chi ^{2}+h_{xx}e^{\mu _{x}{}\chi
}dx^{2}+h_{yy}e^{\mu _{y}{}\chi }dy^{2}\right) +h_{ac}\left( e^{\theta \chi
}\right) ^{c}{}_{b}\omega ^{-\frac{2}{D-3}}dx^{a}dx^{b}\text{ },  \label{sar}
\end{equation}%
whose parameters satisfy 
\begin{eqnarray}
\mu _{x}+\mu _{y}+\theta ^{a}{}_{a} &=&0\text{ },  \notag \\
h_{ac}\theta ^{c}{}_{b} &=&h_{bc}\theta ^{c}{}_{a}\text{ .}  \label{abb}
\end{eqnarray}

As in Class 1, the explicit form of these solutions is obtained by solving
Eqs. (\ref{cs2}), and (\ref{sq}), together with the corresponding coordinate
transformation. For both classes of solutions, as in the vacuum case, $%
h_{ab}=h_{ba}$. Moreover, $h_{xx},$ $h_{yy},$ $h_{zz},$ and the non-zero
values of $h_{ab}$ can be chosen to take the values $\pm 1$.

\section{$\mathcal{N}=2$, $D=4$ Supergravity}
The bosonic sector of ungauged $\mathcal{N}=2$, $D=4$ supergravity coupled
to vector multiplets, with the hypermultiplets consistently truncated, is
described by the Lagrangian 
\begin{equation}
\mathcal{L}_{4}=\omega \left[ R-2g_{A\bar{B}}\,\partial _{\mu }z^{A}\partial
^{\mu }\bar{z}^{B}-\frac{\varepsilon }{2}\left( \operatorname{Im}\,\mathcal{N}
_{IJ}\,\mathcal{F}^{I}\cdot \mathcal{F}^{J}+\operatorname{Re}\,\mathcal{N}_{IJ}\,%
\mathcal{F}^{I}\cdot \tilde{\mathcal{F}}^{J}\right) \right] \text{ .}
\label{Action}
\end{equation}%
The theory contains $n+1$ Abelian gauge fields $\mathcal{A}^{I},$ with field
strengths $\mathcal{F}^{I}=d\mathcal{A}^{I}$, together with $n$ complex
scalar fields $z^{A}$. The scalar fields parametrize a special K\"{a}hler
manifold \cite{superbook}. A convenient description of special geometry is
provided by a $(2n+2)$-dimensional symplectic bundle over the K\"{a}%
hler-Hodge manifold. The covariantly holomorphic symplectic section is 
\begin{equation}
V=%
\begin{pmatrix}
L^{I} \\ 
M_{I}%
\end{pmatrix}
,\quad I=0,\ldots ,n\text{ },
\end{equation}%
where 
\begin{equation}
M_{I}=\mathcal{N}_{IJ}L^{J}\text{ }.
\end{equation}%
This section satisfies 
\begin{equation}
\mathcal{D}_{\bar{A}}V=\left( \partial _{\bar{A}}-\tfrac{1}{2}\partial _{%
\bar{A}}K\right) V=0\text{ },
\end{equation}%
together with the symplectic constraint 
\begin{equation}
i\langle V,\bar{V}\rangle =i\left( \bar{L}^{I}M_{I}-L^{I}\bar{M}_{I}\right)
=1\text{ }.
\end{equation}%
Introducing the holomorphic symplectic section%
\begin{equation}
\Omega =e^{-K/2}V=%
\begin{pmatrix}
X^{I} \\ 
F_{I}%
\end{pmatrix}%
\text{ },\quad 
\end{equation}%
which satisfies 
\begin{equation}
\partial _{\bar{A}}\Omega =0\text{ , }
\end{equation}%
the K\"{a}hler potential is given by 
\begin{equation}
e^{-K}=i\left( \bar{X}^{I}F_{I}-X^{I}\bar{F}_{I}\right) \text{ },
\end{equation}%
while the scalar metric takes the form%
\begin{equation}
g_{A\bar{B}}=\partial _{A}\partial _{\bar{B}}K\text{ }.
\end{equation}

The theory can be formulated in terms of a homogeneous holomorphic
prepotential $F(X^{I})$ of degree two, with $F_{I}=\frac{\partial F}{%
\partial X^{I}}$. Homogeneity implies 
\begin{align}
F& =\frac{1}{2}F_{I}X^{I}\text{ },\quad F_{I}=F_{IJ}X^{J}\text{ },  \notag \\
X^{I}F_{IJK}& =0\text{ },\quad F_{I}\partial _{\mu }X^{I}-X^{I}\partial
_{\mu }F_{I}=0\text{ },
\end{align}%
where 
\begin{equation}
F_{IJ}=\frac{\partial ^{2}F}{\partial X^{I}\partial X^{J}}\text{ },\quad
F_{IJK}=\frac{\partial ^{3}F}{\partial X^{I}\partial X^{J}\partial X^{K}}%
\text{ }.
\end{equation}

Four-dimensional $\mathcal{N}=2$ supergravity admits formulations in all
possible spacetime signatures \cite{m1,m2,m3,m4,m5}. The bosonic sectors of
these theories were constructed in \cite{sig} by dimensional reduction of
the eleven-dimensional supergravity theories of \cite{Hull} on Calabi--Yau
threefolds, followed by further reduction over either spacelike or timelike
circles.

Four-dimensional $\mathcal{N}=2$ supergravity theories coupled to vector and
hypermultiplets in signatures $(0,4)$, $(1,3),$ and $(2,2)$ also arise from
compactifications of type-II string theories with signatures $(0,10)$, $%
(1,9),$ and $(2,8),$ respectively, on Calabi-Yau threefolds \cite{mg}. Here,
we use the notation $(t,s),$ where $t$ and $s$ denote the numbers of
timelike and spacelike dimensions, respectively. The special geometry of
theories with Euclidean and neutral signatures can be formulated by
replacing complex fields with para-complex fields. A unified treatment of
all signatures is achieved by introducing the unit $i_{\epsilon }$,
satisfying $i_{\epsilon }^{2}=\epsilon $, where $\epsilon =-1$ corresponds
to ordinary complex geometry, while $\epsilon =1$ corresponds to
para-complex geometry.

For later use, we summarize several identities of special K\"{a}hler
geometry. The scalar kinetic term can be written as 
\begin{equation}
g_{A\bar{B}}\,\partial _{\mu }z^{A}\partial ^{\mu }\bar{z}%
^{B}=Q_{IJ}\,\partial _{\mu }X^{I}\partial ^{\mu }\bar{X}^{J}\text{ },
\end{equation}%
where 
\begin{align}
\mathcal{N}_{IJ}& =\bar{F}_{IJ}-\epsilon \frac{i_{\epsilon }(NX)_{I}(NX)_{J}%
}{XNX}\text{ },  \notag \\
Q_{IJ}& =e^{K}N_{IJ}+e^{2K}(N\bar{X})_{I}(NX)_{J}\text{ }.  \label{moh}
\end{align}%
Here, 
\begin{equation}
N_{IJ}=i_{\epsilon }(\bar{F}_{IJ}-F_{IJ})\text{ },\quad (NX)_{I}=N_{IJ}X^{J}%
\text{ },\quad (N\bar{X})_{I}=N_{IJ}\bar{X}^{J}\text{ },\quad
XNX=X^{I}N_{IJ}X^{J}\text{ }.
\end{equation}%
These identities will be used extensively in constructing the one-variable
solutions presented below.

\subsection{Class 1 $\mathcal{N}=2,$ $D=4$ supergravity solutions}

We now apply the general results of Section 2 to construct Class 1
one-variable solutions of \ ungauged $\mathcal{N}=2$, $D=4$ supergravity.
Restricting to configurations for which the sections $X^{I}$ are either real
or purely imaginary, the topological term in the Lagrangian (\ref{Action})
does not contribute to the equations of motion. Consequently, the general
formalism developed in the previous section applies directly.

For the metric ansatz (\ref{f1}), the spacetime metric and gauge fields are
given by 
\begin{align}
ds^{2}& =H^{2}\left( \epsilon _{0}d\sigma ^{2}+h_{ac}\left( e^{\left( 2\log
\sigma \right) \lambda }\right) ^{c}{}_{b}dx^{a}dx^{b}\right) +h_{zz}\sigma
^{2p}H^{-2}dz^{2}\text{ },  \notag \\
\operatorname{Im}\mathcal{N}_{IJ}\mathcal{F}^{J\sigma z}{}& =\frac{q_{I}}{\sigma
H^{2}}\text{ }.  \label{fc1}
\end{align}
To solve Eqs. (\ref{g}), and (\ref{pi}) for real sections $X^{I}$, we
introduce the coordinate transformation (\ref{cs}) and adopt the ansatz 
\begin{equation}
H^{2}=4i_{\epsilon }F\text{ .}
\end{equation}%
Using the identities of special K\"{a}hler geometry summarized above, we
obtain 
\begin{equation}
F_{I}=\frac{1}{2}\epsilon i_{\epsilon }\left( A_{I}\sigma ^{p-s}+B_{I}\sigma
^{p+s}\right) \allowbreak \text{ },
\end{equation}%
where $A_{I},$ $B_{I}$, and $s$ are constants. In addition to (\ref{c1}),
the parameters of the solution satisfy 
\begin{equation}
\lambda ^{a}{}_{b}\lambda ^{b}{}_{a}-1=p^{2}-2s^{2}\text{ ,}  \label{lon}
\end{equation}%
together with 
\begin{equation}
{\operatorname{Im}\mathcal{N}^{IJ}}S_{IJ}=\partial _{A}{\operatorname{Im}\mathcal{N}^{IJ}}%
S_{IJ}=0\text{ },
\end{equation}%
where 
\begin{equation}
S_{IJ}=\left( s^{2}A_{I}B_{J}-\frac{1}{4}\epsilon \varepsilon
h_{zz}q_{I}q_{J}\right) \text{ .}
\end{equation}

A second family of solutions is obtained from the metric (\ref{chad}), 
\begin{equation}
ds^{2}=\omega ^{2}\left( \epsilon _{0}d\chi ^{2}+h_{ac}\left( e^{\theta \chi
}\right) ^{c}{}_{b}dx^{a}dx^{b}\right) +h_{zz}e^{\mu \chi }\omega ^{-2}dz^{2}%
\text{ }.  \label{chad2}
\end{equation}%
As before, we restrict to real sections $X^{I}$ and choose 
\begin{equation}
\omega ^{2}=4i_{\epsilon }F\text{ }.
\end{equation}%
The field equations then imply 
\begin{equation}
F_{I}=\frac{1}{2}\epsilon i_{\epsilon }\left( C_{I}e^{\frac{1}{2}\left( \mu
-s\right) \chi }+D_{I}e^{\frac{1}{2}\left( \mu +s\right) \chi }\right) \text{
}\allowbreak .
\end{equation}%
In addition to (\ref{c2}), the metric parameters satisfy 
\begin{equation}
\text{\ }\theta ^{a}{}_{b}\theta ^{b}{}_{a}=\mu ^{2}-2s^{2}\text{ },
\label{lon2}
\end{equation}%
together with 
\begin{equation}
{\operatorname{Im}\mathcal{N}^{IJ}}K_{IJ}=\partial _{A}{\operatorname{Im}\mathcal{N}^{IJ}}%
K_{IJ}=0\text{ },\ 
\end{equation}%
where 
\begin{equation}
K_{IJ}=\left( s^{2}C_{I}D_{J}-\epsilon \varepsilon h_{zz}q_{I}q_{J}\right) 
\text{ }.
\end{equation}

The two families constitute the complete set of Class 1 one-variable
solutions of ungauged $\mathcal{N}=2,$ $D=4$ supergravity within the present
ansatz. Their explicit form depends on the choice of prepotential and the
associated special K\"{a}hler geometry.

\subsection{Class 2 $\mathcal{N}=2,$ $D=4$ supergravity solutions}

We next consider the second class of one-variable solutions, characterized
by gauge fields satisfying (\ref{gog}). The corresponding metric follows
from the general Class 2 solution (\ref{genp}) and is given by 
\begin{equation}
ds^{2}=H^{2}\left( \epsilon _{0}d\sigma ^{2}+h_{xx}\sigma
^{2n_{x}}dx^{2}+h_{yy}\sigma ^{2n_{y}}dy^{2}\right) +h_{zz}\sigma
^{2n_{z}}H^{-2}dz^{2}.
\end{equation}%
The exponents satisfy the constraint 
\begin{equation}
n_{x}+n_{y}+n_{z}=1\ \text{.}  \label{k1}
\end{equation}%
To solve Eqs. (\ref{cs2}), and (\ref{sq}), we restrict to purely imaginary
sections $X^{I}$ and adopt the ansatz 
\begin{equation}
H^{2}=-4i_{\epsilon }F\text{ }.
\end{equation}%
Using the identities of special K\"{a}hler geometry summarized above, we
obtain 
\begin{equation}
X^{I}=\frac{i_{\epsilon }}{2}\left( E^{I}\sigma ^{n_{z}-s}+F^{I}\sigma
^{n_{z}+s}\right) ,
\end{equation}%
where $E^{I},$ $F^{I},$ and $s$ are constants. The parameters satisfy 
\begin{align}
n_{x}^{2}+n_{y}^{2}-n_{z}^{2}-1+2s^{2}& =0\text{ },  \notag \\
\operatorname{Im}\mathcal{N}_{IJ}\left( s^{2}E^{I}F^{J}-\frac{1}{4}\varepsilon
\epsilon _{0}h_{xx}h_{yy}p^{J}p^{I}\right) & =0\text{ },  \notag \\
\partial _{A}\operatorname{Im}\mathcal{N}_{IJ}\left( s^{2}E^{I}F^{J}-\frac{1}{4}%
\varepsilon \epsilon _{0}h_{xx}h_{yy}p^{J}p^{I}\right) & =0\text{ }.
\end{align}

A second family of solutions follows from the metric (\ref{sar}), 
\begin{equation}
ds^{2}=\omega ^{2}\left( \epsilon _{0}d\chi ^{2}+h_{xx}e^{\mu _{x}\chi
}dx^{2}+h_{yy}e^{\mu _{y}\chi }dy^{2}\right) +h_{zz}e^{\mu _{z}\chi }\omega
^{-2}dz^{2},
\end{equation}%
where the constant metric parameters satisfy 
\begin{equation}
\mu _{x}+\mu _{y}+\mu _{z}=0\text{ }.
\end{equation}%
The scalar fields are given by%
\begin{equation}
X^{I}=\frac{i_{\epsilon }}{2}\left( G^{I}e^{\frac{1}{2}\left( \mu
_{z}-s\right) \chi }+H^{I}e^{\frac{1}{2}\left( \mu _{z}+s\right) \chi
}\right) \allowbreak \text{ },
\end{equation}%
where $G^{I},$ $H^{I},$ and $s$ are constants. The remaining constraints are 
\begin{align}
\mu _{x}^{2}+\mu _{y}^{2}-\mu _{z}^{2}+2s^{2}& =0\text{ },  \notag \\
\operatorname{Im}\mathcal{N}{_{IJ}}\left( s^{2}G^{I}H^{J}-\varepsilon \epsilon
_{0}p^{J}p^{I}h_{xx}h_{yy}\right) & =0\text{ },  \notag \\
\partial _{A}\operatorname{Im}\mathcal{N}{_{IJ}}\left( s^{2}G^{I}H^{J}-\varepsilon
\epsilon _{0}p^{J}p^{I}h_{xx}h_{yy}\right) & =0\text{ .}
\end{align}

These two families constitute the complete set of Class 2 one-variable
solutions of ungauged $\mathcal{N}=2,$ $D=4$ supergravity within the present
ansatz. As in the Class 1 case, their explicit realization depends on the
underlying special K\"{a}hler geometry.

\subsection{Explicit Solutions}
As an illustration of the general construction, we present explicit
one-variable solutions for the truncated $\mathcal{N}=8$ supergravity model
considered in \cite{duffliu}, with vanishing scalar potential. This theory
is equivalent to an $\mathcal{N}=2$ supergravity model with prepotential $%
F=-i\sqrt{X^{1}X^{2}X^{3}X^{4}}$ and real sections \cite{silke}$.$
Consequently, we restrict our attention to solutions belonging to the Class
1 family. The spacetime metric is given by the first equation in (\ref{fc1})
with the parameters satisfying (\ref{c1}) and (\ref{lon}), and 
\begin{equation}
H^{2}=4\sigma ^{2p-2s}\prod_{I=1}^{4}\left( 1+\frac{1}{4s^{2}}%
h_{zz}q_{I}^{2}\sigma ^{2s}\right) ^{1/2}.
\end{equation}%
The corresponding gauge field strengths and scalar fields are 
\begin{equation}
\mathcal{F}^{I\sigma z}{}=-q_{I}\frac{\sigma ^{2s-2p-1}}{2\left( 1+\frac{1}{
4s^{2}}h_{zz}q_{I}^{2}\sigma ^{2s}\right) ^{2}}\text{ },\text{ \ \ }X^{I}=
\frac{\sigma ^{p-s}}{\left( 1+\frac{1}{4s^{2}}
h_{zz}q_{I}^{2}\sigma ^{2s}\right)}\prod_{J=1}^{4}\left( 1+\frac{1}{4s^{2}}
h_{zz}q_{J}^{2}\sigma ^{2s}\right) ^{1/2}\text{ }.
\end{equation}

A second family of solutions is described by the spacetime metric (\ref%
{chad2}) with parameters satisfying (\ref{c2}) and (\ref{lon2}), and 
\begin{equation}
\omega ^{2}=4e^{\left( \mu -s\right) \chi }\prod_{I=1}^{4}\left( 1+
\frac{1}{s^{2}}h_{zz}q_{I}^{2}e^{s\chi }\right) ^{1/2}\text{ }.
\end{equation}%
The corresponding gauge field strengths and scalar fields are 
\begin{equation}
\mathcal{F}^{I\chi z}{}=-\frac{q_{I}}{2}\frac{e^{\left( s-\mu \right) \chi }
}{\left( 1+\frac{1}{s^{2}}h_{zz}q_{I}^{2}e^{s\chi }\right) ^{2}}\text{ },%
\text{ \ \ \ }X^{I}=\frac{e^{\frac{1}{2}\left( \mu -s\right) \chi }}
{\left( 1+\frac{1}{s^{2}}h_{zz}q_{I}^{2}e^{s\chi }\right) }\prod_{J=1}^{4}\left( 1+\frac{1}{s^{2}}h_{zz}q_{J}^{2}e^{s\chi
}\right) ^{1/2}%
\allowbreak \text{ }.
\end{equation}

In both families, the constants appearing in the metric must be chosen
appropriately to ensure that the resulting spacetime has Lorentzian
signature.

\section{$SL(N,\mathbb{R})/SO(N,\mathbb{R})$ coset models and explicit
solutions}
In this section, we apply the general formalism developed in Section 2 to
gravitational theories in which the scalar fields parametrize the symmetric
coset manifold $SL(N,\mathbb{R})/SO(N,\mathbb{R})$. These theories are
described by the Lagrangian 
\begin{equation}
\mathcal{L}_{D}=\omega \left( R-{{\frac{{1}}{{2}}}}\sum_{A=1}^{N-1}\partial
_{\mu }\phi ^{A}\partial ^{\mu }\phi ^{A}-{\frac{1}{4}}\sum_{I=1}^{N}\frac{1%
}{(X^{I})^{2}}\mathcal{F}_{\mu \nu }^{I}\mathcal{F}^{I\mu \nu \,}\right) 
\text{ }.  \label{dd}
\end{equation}
The $N$ functions $X^{I}$ are determined by the independent $(N-1)$
dilatonic fields $\phi ^{A}$ and satisfy the constraint 
\begin{equation*}
\prod_{I=1}^{N}X^{I}=1\,.
\end{equation*}
They may be parametrized as 
\begin{equation}
X^{I}=e^{-{\frac{1}{2}}B^{I}.\text{ }\phi }\text{ },  \label{ddxdef}
\end{equation}%
where 
\begin{equation}
B^{I}.\text{ }\phi =\sum_{A}B_{A}^{I}\text{ }.\text{ }\phi ^{A}\text{ }.
\end{equation}%
The vectors $B^{I}$ are the weight vectors of the fundamental representation
of $SL(N,\mathbb{R})$ and satisfy 
\begin{equation}
B^{I}\text{ }.\text{ }B^{J}=8\delta ^{IJ}-{\frac{8}{N}}\,\text{\ },\qquad
\sum_{I}B^{I}=0\,\text{\ }.  \label{dotprod}
\end{equation}

Applying the general analysis of Section 2, one finds that consistent
one-variable solutions exist only when the spacetime dimension $D$ and the
parameter $N$ satisfy 
\begin{equation}
D=\frac{3N-8}{N-4}\text{.}  \label{con}
\end{equation}%
This condition admits only the three possibilities $(N,D)=(8,4),(6,5),(5,7)$%
. Remarkably, the same condition also arises in the construction of domain
wall and charged S-brane solutions in the gauged versions of these theories 
\cite{Cv, Gutsabra}.

\subsection{First family}
The first family is described by the metric (\ref{f1}), with 
\begin{equation}
H^{N(D-3)}=\prod_{I=1}^{N}\left( \sigma ^{p-s}+\frac{N(D-3)}{2s^{2}\left(
D-2\right) }h_{zz}q_{I}^{2}\sigma ^{p+s}\right) \text{ }.
\end{equation}%
The corresponding scalar fields and gauge field strengths are 
\begin{eqnarray}
X^{I} &=&H^{(D-3)}\left( \sigma ^{p-s}+\frac{N(D-3)}{2s^{2}\left( D-2\right) 
}h_{zz}q_{I}^{2}\sigma ^{p+s}\right) ^{-1}\text{ },\text{ \ \ } \\
\mathcal{F}^{I\sigma z} &=&\frac{2q_{I}}{\sigma }H^{2(D-4)}\left( \sigma
^{p-s}+\frac{N(D-3)}{2s^{2}\left( D-2\right) }h_{zz}q_{I}^{2}\sigma
^{p+s}\right) ^{-2}\text{ }.
\end{eqnarray}%
In addition to (\ref{c1}), the parameters satisfy the constraint 
\begin{equation}
\lambda ^{a}{}_{b}\lambda ^{b}{}_{a}-1=-\frac{1}{4}\left(
3p^{2}+s^{2}\right) \text{ }.
\end{equation}
\subsection{Second family}
The second family is described by the metric (\ref{chad}), where 
\begin{equation}
\omega ^{N(D-3)}=\prod_{I=1}^{N}\left( e^{\frac{1}{2}(\mu -s)\chi }+\frac{%
2N(D-3)}{s^{2}\left( D-2\right) }h_{zz}q_{I}^{2}e^{\frac{1}{2}(\mu +s)\chi
}\right) \text{ }.
\end{equation}%
In addition to (\ref{c2}), the parameters of the metric satisfy 
\begin{equation}
\theta _{\text{ }b}^{a}\theta _{\text{ }a}^{b}=-\frac{1}{4}\left( 3\mu
^{2}+s^{2}\right) \text{ }.
\end{equation}%
The corresponding scalar fields and gauge field strengths are 
\begin{align}
\omega ^{-(D-3)}X^{I}& =\left( e^{\frac{1}{2}(\mu -s)\chi }+\frac{2N(D-3)}{%
s^{2}\left( D-2\right) }h_{zz}q_{I}^{2}e^{\frac{1}{2}(\mu +s)\chi }\right)
^{-1}\text{ },  \notag \\
\mathcal{F}^{I\chi z}& =2q_{I}\omega ^{2(D-4)}\left( e^{\frac{1}{2}(\mu
-s)\chi }+\frac{2N(D-3)}{s^{2}\left( D-2\right) }h_{zz}q_{I}^{2}e^{\frac{1}{2%
}(\mu +s)\chi }\right) ^{-2}\text{ }.
\end{align}
\subsection{Third family}
The third family is described by the metric (\ref{genp}), with 
\begin{equation}
H^{N}=\prod_{I=1}^{N}\left( \sigma ^{\lambda ^{a}{}_{a}-s}-\frac{N(D-3)}{%
8s^{2}\left( D-2\right) }\epsilon _{0}h_{xx}h_{yy}\left( p^{I}\right)
^{2}\sigma ^{\lambda ^{a}{}_{a}+s}\right) \text{ }.
\end{equation}%
The corresponding scalar fields and gauge field strengths are 
\begin{equation}
X^{I}=\frac{1}{H}\left( \sigma ^{\lambda ^{a}{}_{a}-s}-\frac{N(D-3)}{%
8s^{2}\left( D-2\right) }\epsilon _{0}h_{xx}h_{yy}\left( p^{I}\right)
^{2}\sigma ^{\lambda ^{a}{}_{a}+s}\right) \text{ },\text{ \ \ \ }\mathcal{F}%
_{xy}^{I}=p^{I}\text{ }.
\end{equation}%
In addition to (\ref{ab}), the parameters satisfy 
\begin{equation}
n_{x}^{2}+n_{y}^{2}+\lambda ^{a}{}_{b}\lambda ^{b}{}_{a}-1=\frac{1}{4}\left(
\left( \lambda ^{a}{}_{a}\right) ^{2}-s^{2}\right) \text{ }.
\end{equation}
\subsection{Fourth family}
The fourth family is described by the metric (\ref{sar}), where 
\begin{equation}
\omega =\prod_{I=1}^{N}\left( e^{\frac{1}{2}\left( \theta
^{a}{}_{a}-s\right) \chi }-\frac{N(D-3)}{2s^{2}\left( D-2\right) }\epsilon
_{0}h_{xx}h_{yy}\left( p^{I}\right) ^{2}e^{\frac{1}{2}\left( \theta
^{a}{}_{a}+s\right) \chi }\right) ^{\frac{1}{N}}.
\end{equation}%
The corresponding scalar fields and gauge field strengths are 
\begin{equation}
X^{I}=\frac{1}{\omega }\left( e^{\frac{1}{2}\left( \theta
^{a}{}_{a}-s\right) \chi }-\frac{N(D-3)}{2s^{2}\left( D-2\right) }\epsilon
_{0}h_{xx}h_{yy}\left( p^{I}\right) ^{2}e^{\frac{1}{2}\left( \theta
^{a}{}_{a}+s\right) \chi }\right) \text{ },\text{ \ \ }\mathcal{F}%
_{xy}^{I}=p^{I}\text{ }.
\end{equation}%
In addition to (\ref{abb}), the parameters satisfy 
\begin{equation}
\mu _{x}^{2}+\mu _{y}^{2}+\theta ^{a}{}_{b}\theta ^{b}{}_{a}=\frac{1}{4}%
\left( \left( \theta ^{a}{}_{a}\right) ^{2}-s^{2}\right) \text{ }.
\end{equation}

These four families provide explicit one-variable solutions for
gravitational theories whose scalar fields parametrize the symmetric coset
spaces $SL(N,\mathbb{R})/SO(N,\mathbb{R})$. The consistency conditions
restrict the existence of these solutions to the three cases $%
(N,D)=(8,4),(6,5),(5,7)$.

\section{Summary}
In this paper, we have constructed four families of one-variable solutions
for a broad class of $D$-dimensional gravitational theories coupled to
scalar and Abelian gauge fields. The resulting spacetimes are described by
metrics depending on a single coordinate, with their structure determined by
a metric function and constant matrices subject to a set of algebraic
constraints.

We applied the general formalism to ungauged $\mathcal{N}=2$, $D=4$
supergravity coupled to vector multiplets and obtained four families of
solutions supported by nontrivial scalar and gauge field configurations. As
an explicit illustration, we constructed solutions for a truncation of $%
\mathcal{N}=8,$ $D=4$ supergravity, demonstrating how the general framework
yields closed-form solutions in a physically relevant model. We also derived
explicit one-variable solutions for theories whose scalar fields parametrize
the symmetric coset manifolds $SL(N,\mathbb{R})/SO(N,\mathbb{R})$. In this
case, the consistency conditions restrict the existence of such solutions to
the three admissible pairs $(N,D)=(8,4)$, $(6,5)$, $(5,7$).

The framework developed in this work is sufficiently general to be applied
to a wide class of gravitational theories involving scalar and Abelian gauge
fields. Several directions remain open for future investigation. These
include constructing explicit solutions for more general ungauged $\mathcal{N%
}=2$, $D=4$ supergravity models, extending the analysis to gauged
supergravity theories in which scalar potentials may give rise to new
classes of one-variable geometries, and undertaking a detailed study of the
global, causal, and thermodynamic properties of the resulting spacetimes. It
would also be of interest to investigate the higher-dimensional origin of
these solutions within string theory and M-theory compactifications, as well
as their potential applications in cosmology.

\end{document}